\documentclass[preprints,article,submit,accept,moreauthors]{Definitions/mdpi} 

\firstpage{1} 
\pubvolume{1}
\issuenum{1}
\articlenumber{0}
\pubyear{2023}
\copyrightyear{2023}
\datereceived{ } 
\daterevised{ } 
\dateaccepted{ } 
\datepublished{ } 
\hreflink{https://doi.org/} 
\usepackage{amsfonts}
\usepackage{amssymb}
\usepackage{algpseudocode}
\usepackage{algorithm}
\usepackage{booktabs}
\usepackage{tikz}
\usepackage{tikz-qtree}
\usepackage[edges]{forest}
\usepackage{makecell}
\usepackage{subcaption}
\usepackage{float}

\Title{Phylogenetic tree distance computation over succinct representations}
\TitleCitation{Phylogenetic tree distance computation over succinct representations}
\Author{António Pedro Branco $^{1,2}$, Cátia Vaz $^{1,3*}$ and Alexandre P. Francisco $^{1,2}$}
\AuthorNames{António Pedro Branco, Cátia Vaz and Alexandre P. Francisco}
\AuthorCitation{Branco, A. P.; Vaz, C.; Francisco, A. P.}
\address{%
$^{1}$ \quad
Instituto de Engenharia de Sistemas e Computadores, I\&D em Lisboa (INESC-ID), Portugal\\
$^{2}$ \quad Instituto Superior Técnico, Universidade de Lisboa, Portugal\\
$^{3}$ \quad Instituto Superior de Engenharia de Lisboa, Instituto Politécnico de Lisboa, Portugal\\
}

\corres{cvaz@cc.isel.ipl.pt}

\renewcommand{\contentleftcolumn}{}

\abstract{There are several tools available to infer phylogenetic trees,
which depict the evolutionary relationships among biological entities such as viral and bacterial strains in infectious outbreaks, or cancerous cells in tumor progression trees.
These tools rely on several inference methods available to produce phylogenetic trees, with resulting trees not being unique. Thus, methods for comparing phylogenies that are capable of revealing where two phylogenetic trees agree or differ are required.
An approach is then to compute a similarity or dissimilarity measure between trees, with the Robinson-Foulds distance being one of the most used, and which can be computed in linear time and space. Nevertheless, given the large and increasing volume of phylogenetic data, phylogenetic trees are becoming very large with hundreds of thousands of leafs. In this context,
space requirements become an issue both while computing tree distances and while storing trees.
We propose then an efficient implementation of the Robinson-Foulds distance over trees succinct representations.
Our implementation generalizes also the Robinson-Foulds distances to labelled phylogenetic trees, {\em i.e.}, trees containing labels on all nodes, instead of only on leaves. Experimental results show that we are able to still achieve linear time while requiring less space. Our implementation is available as an open-source tool at
\url{https://github.com/pedroparedesbranco/TreeDiff}.}


\begin{document}

\section{Introduction}
Comparative evaluation of differences, similarities, and distances between phylogenetic trees is a fundamental task in computational phylogenetics~\cite{felsenstein2004inferring}. 
There are several measures for assessing differences between two phylogenetic trees, some of them based on rearrangements, others based on topology dissimilarity and in this case some of them take also into account the branch-length~\cite{kuhner2015practical}. The rearrangement measures are based on finding the minimum number of rearrangement steps required to transform one tree into the other. Possible rearrangement steps include nearest-neighbor interchange (NNI), subtree pruning and regrafting (SPR), and tree bisection and reconnection (TBR). Unfortunately such measures are seldom used in practice for large studies as they are expensive to calculate in general. NP-completeness has been shown for distances based on NNI~\cite{li1999twist}, TBR~\cite{allen2001subtree}, and SPR~\cite{bordewich2005computational}.
Measures based on topological dissimilarity are commonly used. One of the most used is the topological distance of Robinson and Foulds~\cite{robinson1981comparison} (RF) and its branch-length variation~\cite{robinson2006comparison}. The RF-like distances, when applied to rooted trees are all based on the idea of shared clades, i.e., specific kinds of subtrees (for rooted trees) or branches defined by possession of exactly the same tips (the leaves of the tree). Namely, it quantifies the dissimilarity between two trees based on the differences in their clades, for rooted trees, or bipartitions if applied to unrooted trees.
Formally, a clade or cluster $C(n)$ of a tree $T$ is defined by the set of leaves (or nodes in fully labelled trees) that are a descendant from a particular internal node $n$.

The computation of the RF distance can be achieved in linear time and space. A linear time algorithm for calculating the RF distance was first proposed by Day~\cite{dayalgorithm}. It efficiently determines the number of bipartitions or clades that are present in one tree but not in the other. 
Furthermore, there have been advancements in the computation of the RF distance that achieve sublinear time complexity; Pattengale et al.~\cite{pattengale2007efficiently} introduced an algorithm that can compute an approximation of the RF distance in sublinear time. However, it is important to note that the sublinear algorithm is not exact and may introduce some degree of error in the computed distances.
Recently, it was proposed a linear time and space algorithm~\cite{briand2022linear} that addresses the labelled Robinson-Foulds (RF) distance problem, considering labels assigned to both internal and leaf nodes of the trees. However, their distance is based on edit distance operations applied to nodes and thus do not give the exact RF value. Moreover the implementation is based on the Day algorithm~\cite{dayalgorithm}.

Considering however the increasingly large volume of phylogenetic data~\cite{zhou2020enterobase}, the size of phylogenetic trees has grown substantially, often consisting of hundreds of thousands of leaf nodes. This poses significant challenges in terms of space requirements for computing tree distances, or even for storing trees. Hence, we propose an efficient implementation of the Robinson-Foulds distance over succinct representations of trees~\cite{navarro2016compact}. Our implementation not only addresses the standard Robinson-Foulds distance but it also extends it to handle fully labelled and weighted phylogenetic trees. By leveraging succinct representations, we are able to achieve practical linear time while significantly reducing space requirements. Experimental results demonstrate the effectiveness of our implementation in spite of expected trade-offs on running time overhead versus space requirements due to the use of succinct data structures.

\section{Background}
A phylogenetic tree, denoted as $T=(V, E)$ is defined as a connected acyclic graph.  The set $V$, also denoted as $V(T)$ represents the vertices (nodes) of tree $T$. The set $E\subseteq V\times V$, also denoted as $E(T)$ represents the edges (links) of tree $T$. The leaves of the tree, which are the vertices of degree one, are labelled with data that represents species, strains or even genes.
A phylogenetic tree represents then the evolutionary relationships among taxonomical groups, or taxa.

In most phylogenetic inference methods, the internal vertices of the tree represent hypothetical or inferred common ancestors of the entities represented by the leaves. These internal vertices do not typically have a specific label or represent direct data. However, there are some distance based phylogenetic inference methods that infer a fully labelled phylogenetic tree~\cite{vaz2021distance}. In such cases, the internal vertices may also have labels associated with them, providing additional information about the inferred common ancestors. We will denote the labels of a phylogetic tree $T$ by $L(T)$.
The inferred phylogenetic trees can be rooted or unrooted, depending on whether or not a specific root node is assigned. In the rooted ones,  there exists a distinguished vertex called the root of $T$.

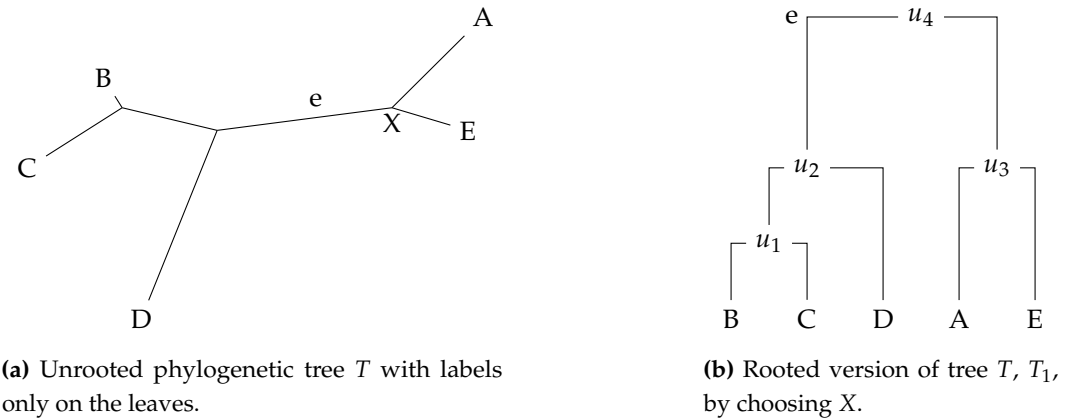
\begin{figure}[H]
    \centering   
    	\subcaptionbox{Unrooted phylogenetic tree $T$ with labels only on the leaves.}{
    		\begin{tikzpicture}
    		\node (a)  at (0,3) {C};
    		\node (b) at (1,4.2) {B};
    		\node (c) at (6,5) {A};
    		\node (d) at (5.8,3.5) {E};
    		\node (e) at (1.5,1) {D};
    		\node (ab) at (1.25,3.8) {};
    		\node (cd) at (4.8,3.8) {};
    		\node (x) at (4.8,3.6) {X};
                \node (edge) at (3.8,3.9) {e};
    		\node (abcd) at (2.5,3.5) {};
    		\draw (a)-\\(ab.center);
    		\draw (b)-\\(ab.center);
    		\draw (c)-\\(cd.center);
    		\draw (d)-\\(cd.center);
    		\draw (ab.center)-\\(abcd.center);
    		\draw (cd.center)-\\(abcd.center);
    		\draw (e)-\\(abcd.center);
    		\end{tikzpicture}
    }
  \hfill
      \subcaptionbox{Rooted version of tree $T$, $T_1$, by choosing $X$.}{
    		\begin{tikzpicture}
			\node (b) at (1,0) {B};
			\node (c) at (2,0) {C};
			\node (d) at (3,0) {D};
			\node (a) at (4,0) {A};
			\node (e) at (5,0) {E};
			\node (bc) at (1.5,1) {$u_1$};
			\node (bcd) at (2,2) {$u_2$};
			\node (ae) at (4.5,2) {$u_3$};
			\node (all) at (3.5,4) {$u_4$};
                \node (edge) at (1.8,4) {e};
			\draw  (b) |- (bc);
			\draw  (c) |- (bc);
			\draw  (d) |- (bcd);
			\draw  (a) |- (ae);
			\draw  (e) |- (ae);
			\draw  (bc) |- (bcd);
			\draw  (bcd) |- (all);
			\draw  (ae) |- (all);
                \end{tikzpicture}
                \label{tree_X} 
    }
    \caption{An unrooted phylogenetic tree (a) and a rooted phylogenetic tree (b) that resulted from transforming the unrooted one by selecting node $X$ as root; $e$ denotes the equivalent edge in both trees. The set of labels of $T_1$, $L(T_1)$ is $\{A,B,C,D,E\}$.} 
    \label{fig:Treesexample1} 
\end{figure}

To calculate the Robinson-Foulds (RF) distance on a rooted or unrooted tree, you need to compare the clades or bipartitions of the trees, respectively. As mentioned before, a {\em clade} is a group on a tree that includes a node and all of the lineages descended from that node. Thus, clades can be seen as the {\em subtrees} of a phylogenetic tree. A {\em bipartition} of a tree is induced by an edge removal ~\cite{felsenstein2004inferring}. Nevertheless, we can convert an unrooted tree into rooted one by adding an arbitrary root node~\cite{huson2010phylogenetic} (see Figure~\ref{fig:Treesexample1}). A measure based on Robison-Foulds distance to compare unrooted with rooted trees was also proposed~\cite{gorecki2012robinson}.
In this work, we consider rooted trees.
Let $T$ be a rooted tree. A {\em sub-tree} $S$ of $T$, which we will denote by $S(T)$ is a tree such that $V(S) \subseteq V(T)$, $E(S) \subseteq E(T)$, and the endpoints of any edge in $E(S)$ must belong to $V(S)$.  We will denote by $T(v)$ the {\em maximum sub-tree} of $T$ that is rooted at $v$. Thus, for each $v \in V(T)$ we define the {\em cluster} at $v$ in $T$ as $c(v)=L(T(v))$. For instance, the rooted tree in Figure~\ref{fig:Treesexample1} contains 9 clusters, namely: $c(B)=\{B\}$, $c(C)=\{C\}$, $c(D)=\{D\}$, $c(A)=\{A\}$, $c(E)=\{E\}$, $c(u_1)=\{B,C\}$, $c(u_2)=\{B,C,D\}$, $c(u_3)=\{A,E\}$ and $c(u_4)=\{B,C,D,A,E\}$, with $u_i$ denoting unnamed or unlabelled nodes in $T$. The set of all clusters present in $T$ is denoted by $C(T)$, i.e., $C(T)=\cup_{v \in V(T)} \{c(v)\}$. In Figure~\ref{fig:Treesexample1}, $C(T_1)$ contains the previous 9 clusters.
The Robison-Foulds distance measures the dissimilarity between two trees by counting the differences in the clades they contain if rooted.
\begin{Definition}\label{def1}
The Robinson-Foulds (RF) distance between two rooted trees $T_1$ and $T_2$, with unique labelling in leafs,
is defined by:
$$RF(T_1,T_2)=|C(T_1) \setminus C(T_2)| \cup |C(T_2) \setminus C(T_1)|.$$
\end{Definition}
Notice that this definition is equivalent to $RF(T_1,T_2)=|C(T_1)| + |C(T_2)| - 2 |C(T_1) \cap C(T_2)|$, commonly found in literature. Note also that most software implementations usually divide the RF distance by 2~\cite{kuhner2015practical}, and further normalize it.

When dealing with fully labelled phylogenetic trees, labels are not only present in leaves but also in internal nodes. Figure~\ref{fig:Treesexample3} depicts two rooted and fully labelled phylogenetic trees. The clusters of tree $T_2$ are: $C(A)=\{A\}$, $C(B)=\{B\}$, $C(C)=\{C\}$, $C(D)=\{D\}$, $C(E)=\{E\}$, $C(F)=\{B,C,F\}$, $C(G)=\{B,C,F,D,G\}$, $C(H)=\{A,E,H\}$ and $C(I)=\{B,C,F,D,G,A,E,H,I\}$. The RF distance given in Definition~\ref{def1} can be extended to fully labelled trees by just noting that, given $v\in V(T)$, $c(v)=L(T(v))$ includes now the labels of all nodes in subtree $T(v)$; and hence the clusters in $C(T)$ include the labels for all nodes also.
\begin{figure}[H]
\centering
    \subcaptionbox{A rooted labelled phylogenetic tree $T_2$.}{
\begin{tikzpicture}
			\node (b) at (1,0) {B};
			\node (c) at (2,0) {C};
			\node (d) at (3,0) {D};
			\node (a) at (4,0) {A};
			\node (e) at (5,0) {E};
			\node (bc) at (1.5,1) {F};
			\node (bcd) at (2,2) {G};
			\node (ae) at (4.5,2) {H};
			\node (all) at (3.5,4) {I};
 		\draw  (b) |- (bc);
			\draw  (c) |- (bc);
			\draw  (d) |- (bcd);
			\draw  (a) |- (ae);
			\draw  (e) |- (ae);
			\draw  (bc) |- (bcd);
			\draw  (bcd) |- (all);
			\draw  (ae) |- (all);
                \end{tikzpicture}
                 }
  \hfill
      \subcaptionbox{A rooted labelled phylogenetic tree $T_3$.}{
      \begin{tikzpicture}
			\node (b) at (1,0) {B};
			\node (c) at (2,0) {C};
			\node (d) at (3,0) {D};
			\node (a) at (4,0) {A};
			\node (e) at (5,0) {E};
			\node (bc) at (2.5,1) {F};
			\node (bcd) at (2,2) {G};
			\node (ae) at (4.5,2) {H};
			\node (all) at (3.5,4) {I};
 		\draw  (d) |- (bc);
			\draw  (c) |- (bc);
			\draw  (b) |- (bcd);
			\draw  (a) |- (ae);
			\draw  (e) |- (ae);
			\draw  (bc) |- (bcd);
			\draw  (bcd) |- (all);
			\draw  (ae) |- (all);
                \end{tikzpicture}
                }
                \caption{Two rooted labelled phylogenetic trees $T_2$ and $T_3$. Both have a unique set of labels $\{A,B,C,D,E,F,G,H,I\}$.} 
    \label{fig:Treesexample3} 
\end{figure}
\begin{Definition}\label{def2}
The extended Robinson-Foulds (RF) distance between two fully labelled rooted trees $T_1$ and $T_2$, with unique labelling in all nodes, is defined by:
$$eRF(T_1,T_2)=|C(T_1) \setminus C(T_2)| \cup |C(T_2) \setminus C(T_1)|.$$
\end{Definition}
Let us consider the trees $T_2$ and $T_3$ in Figure~\ref{fig:Treesexample3}; $eRF(T_2, T_3)=2$ since there are only two distinct clusters in $T_2$ and $T_3$, namely $\{B,C,F\}$ and $\{C,D,F\}$.

Even though RF can give a very reliable distance between two trees in terms of their topology, sometimes it is convenient to consider weights. A version of Robinson-Foulds distance for weighted trees (wRF) takes then into account the branch weights of both trees~\cite{robinson1979comparison}. These weights represent for instance the varying levels of correction for DNA sequencing errors.
Let $T$ be a phylogenetic rooted tree, $v \in V(T)$, $w:E(T)\rightarrow \mathbf{R}$, $e\in E(T)$ the edge with target $v$, and $c(v)$ the cluster at $v$ in $T$ as before; the weight of cluster $c(v)$ in $T$, $w_T(c(v))$ is defined as $w(e)$.
\begin{Definition}\label{def3}
  The weighted Robinson-Foulds (wRF) distance between two rooted trees $T_1$ and $T_2$, with unique labelling in leafs, is defined as: $$wRF(T_1,T_2)= \sum_{c \in C(T_1) \cap C(T_2)} | w_{T_{1}}(c) - w_{T_{2}}(c)| + \sum_{c \in C(T_1) \setminus C(T_2)} w_{T_{1}}(c) + \sum_{c \in C(T_2) \setminus C(T_1)} w_{T_{2}}(c).$$
\end{Definition}
In fact, at lower levels of sequencing error, RF distance usually exhibits a consistent pattern, showcasing the superiority of proper correction for both over- and undercorrection for DNA sequencing error. Interestingly, in this context, correction had no impact on the RF scores.
Conversely, at extremely high levels of sequencing error, wRF results suggested that proper correction and overcorrection were equivalent, while the RF scores demonstrated that overcorrection led to the destruction of topological recovery. Then, both measures complement themselves~\cite{kuhner2015practical}. Definition~\ref{def4} extends wRF to labelled rooted phylogenetic trees assuming the extension of $c(v)$ and $C(T)$ as before.
\begin{Definition}\label{def4}
The weighted labelled Robinson-Foulds (weRF) distance between two fully labelled rooted trees $T_1$ and $T_2$, with unique labelling in all nodes, is defined as follows: $$weRF(T_1,T_2)= \sum_{c \in C(T_1) \cap C(T_2)} | w_{T_{1}}(c) - w_{T_{2}}(c)| + \sum_{c \in C(T_1) \setminus C(T_2)} w_{T_{1}}(c) + \sum_{c \in C(T_2) \setminus C(T_1)} w_{T_{2}}(c).$$
\end{Definition}

Phylogenetic trees are commonly represented and stored using the well known newick tree format. For instance, the rooted tree in Figure~\ref{fig:Treesexample1} can be represented as \texttt{(((B, C), D),(A, E));}. In this example we have only labels on the leafs and we do not have edge weights. The newick format supports however internal labelled vertices and edge weights, e.g., \texttt{(((B:0.2, C:0.2)W:0.3, D:0.5)X:0.6, (A:0.5, E:0.5)Y:0.6)Z;}. Thus, it is common that tools that compute the RF distance support as input trees in newick format.
Notice also that some phylogenetic inference algorithms may produce $n$-ary phylogenetic trees, such is the case, for instance, of goeBURST~\cite{goeburst} or MSTree V2~\cite{zhou2018grapetree}.

\section{Approach}
Our approach consists in using succinct data structures, namely encoding each tree using balanced parentheses and represent it as a bit vector. Each tree is then represented by a bit vector of size $2n$ bits, where $n$ is the number of nodes present in the tree. This space is further increased by $o(n)$ bits to support efficiently primitive operations over the bit vector~\cite{10.1145/2601073}. For comparing trees, we need to store also the labels permutation corresponding to each tree. Each permutation can be represented in $(1 + \varepsilon) n\log n$ bits, while answering permutation queries in constant time and inverse permutation queries in $O(1/\varepsilon)$ time~\cite{navarro2016compact}.
Hence, each tree with $n$ nodes can be represented in $2n + o(n) + (1 + \varepsilon) n\log n$ bits.


\subsection{Balanced parentheses representation}
A balanced parentheses representation is a method frequently used to represent hierarchical relationships between nodes in a tree, closely related to the newick format described before. Figure~\ref{fig:tree1} depicts an example of a rooted labelled phylogenetic tree and its balanced parentheses representation. In this kind of representation, there are some key rules to understand how a tree can be represented by a sequence of parentheses.
The first rule is that each node is mapped to a unique \texttt{index} given by the tree pre-order transversal.
For instance, in Figure~\ref{fig:tree1}, node with label $A$ is mapped to $4$. The second rule is that each node is represented by a pair of opening and closing parentheses. For example, in Figure~\ref{fig:tree1} node with \texttt{index} $4$ is represented by the parentheses in $4$th and $5$th positions. In terms of memory representation, the bit vector keeps this information, where the opening parentheses will be represented as a $1$ and the closing ones as a $0$. This is the reason why the size of the bit vector is $2$ times the number of nodes. Notice that it is not necessary to explicitly store the \texttt{index} of each node.
\begin{figure}[H]
  \centering
    \subcaptionbox{A rooted labelled phylogenetic tree $T_4$  with labels on leaves. $u_1$, $u_2$, $u_3$ and $u_4$ are unlabelled nodes.}{
  \scalebox{1}{
\begin{tikzpicture}
			\node (b) at (1,0) {A};
			\node (c) at (2,0) {B};
			\node (d) at (3,0) {C};
			\node (a) at (4,0) {D};
			\node (e) at (5,0) {E};
            \node (k) at (6,0) {F};
			\node (bc) at (1.5,1) {$u_1$};
			\node (bcd) at (2,2) {$u_2$};
			\node (aek) at (5,2) {$u_3$};
			\node (all) at (3.5,4) {$u_4$};
 		\draw  (b) |- (bc);
			\draw  (c) |- (bc);
			\draw  (d) |- (bcd);
			\draw  (a) |- (aek);
			\draw  (e) |- (aek);
            \draw  (k) |- (aek);
			\draw  (bc) |- (bcd);
			\draw  (bcd) |- (all);
			\draw  (aek) |- (all);
                \end{tikzpicture}
              }}
  \hfill
      \subcaptionbox{Index $i$ of each node in $T_4$.}{
  \scalebox{0.8}{
  \begin{forest}
  for tree={
        grow=south,
         minimum size=5ex, inner sep=1pt,
        s sep=7mm
            }
    [1,circle,draw
        [2,circle,draw
            [3,circle,draw,name=cluster
                [4,circle,draw=green, top color=green!5, bottom color=green!30, name=lowestl
                    [\textbf{A},no edge]
                ]
                [5,circle,draw=red, top color=red!5, bottom color=red!30, name=lowestr
                    [\textbf{B},no edge]
                ]
            ]
            [6,circle,draw=yellow, top color=yellow!5, bottom color=yellow!30
                [\textbf{C},no edge]
            ]
        ]
        [7,circle,draw,
            [8,circle,draw=blue, top color=blue!5, bottom color=blue!30
                [\textbf{D},no edge]
            ]
            [9,circle,draw=orange, top color=orange!5, bottom color=orange!30
                [\textbf{E},no edge]
            ]
            [10,circle,draw=brown, top color=brown!5, bottom color=brown!30
                [\textbf{F},no edge]
            ]
        ]
    ]
    \node[draw,fit={(cluster) (lowestl) (lowestr)}] {};
  \end{forest}
}}

\vspace{3ex}

 \subcaptionbox{The balanced representation of $T_4$.}{
   $
   B =
\begin{array}{cccccccccccccccccccc}
( & ( & ( & ( & ) & ( & ) & ) & ( & ) & ) & ( & ( & ) & ( & ) & ( & ) & ) & )\\
1_o & 2_o & 3_o & 4_o & 4_c & 5_o & 5_c & 3_c & 6_o & 6_c & 2_c & 7_o & 8_o & 8_c & 9_o & 9_c & 10_o & 10_c & 7_c & 1_c\\
\end{array}
   $
 }
  \caption{Tree $T_4$ and its balanced parentheses representation. Each pair of parentheses correspond to a unique node \texttt{index} identified as $i$ in the tree.}\label{fig:tree1}
\end{figure}

The third rule is that for all nodes present in the tree, all their descendants appear after the opening parentheses and before the closing parentheses that represent them. For example, in Figure~\ref{fig:tree1}, the opening and closing parentheses that represent the third node are in positions 3 and 8, respectively. Then, we can conclude that the parentheses that represents its descendants (nodes 4 and 5) are in positions 4, 5, 6, and 7, which are between 3 and 8.
Throughout this document we will refer to an \texttt{index} of a node as $i$ and to a \texttt{position} in a bit vector as $p$.

\subsection{Operations} \label{subsec:operations}
The balanced parentheses representation contains several operations that are fundamental to manipulate this structure effectively. The most important ones in the present context are the operations \texttt{Rank1}, \texttt{Rank10}, \texttt{Select1}, \texttt{Select0},  \texttt{FindOpen}, \texttt{FindClose}, and \texttt{Enclose}. Then, we added operations \texttt{PreOrderMap},  \texttt{IsLeaf}, \texttt{LCA}, \texttt{ClusterSize}, \texttt{PreOrder\-Select}, \texttt{PostOrderSelect} and \texttt{NumLeaves}, that mostly use the fundamental operations just mentioned before. In Table~\ref{fig:operations} it is possible to see the list of operations as well as their meaning and their run time complexity; see the text by Navarro~\cite{navarro2016compact} for details on primitive operations. More details in the implementation of the added operations can be found in Appendix~\ref{appendix2}.

\begin{table}[H]
  \caption{Operations to manipulate trees (see~\cite{navarro2016compact} for details).}\label{fig:operations}
\centering
\begin{tabular}{lll}
\toprule
\multicolumn{1}{l}{\textbf{Operation}} & \multicolumn{1}{l}{\textbf{Meaning}} & \multicolumn{1}{l}{\textbf{Complexity}} \\
\midrule
Rank1($bv, p$) &  \makecell[l]{ Given a bit vector $bv$ and a position $p$, returns \\ the number of occurrences of '1' until $p$.} & $O(1)$ \\ \\
Rank10($bv, p$) & \makecell[l]{ Given a bit vector $bv$ and a position $p$, returns \\ the number of occurrences of the sequence '10' \\ until $p$.} & $O(1)$\\ \\
Select1($bv, i$) & \makecell[l]{ Given a bit vector $bv$ and an occurrence $i$, returns \\ the position where the ith one is present.} &$O(1)$\\ \\
Select0($bv, i$) & \makecell[l]{ Given a bit vector $bv$ and an occurrence $i$, returns \\ the position where the ith zero is present.} & $O(1)$\\ \\
FindOpen($bv, p$) & \makecell[l]{Given a bit vector $bv$ and a position $p$, returns \\ the position where the corresponding opening \\ parentheses is located.} & $O(\log(n))$\\ \\
FindClose($bv, p$) & \makecell[l]{Given a bit vector $bv$ and a position $p$, returns \\ the position where the corresponding closing \\ parentheses is located.} & $O(\log(n))$\\ \\
Enclose($bv, p$) & \makecell[l]{Given a bit vector $bv$ and a position $p$, returns \\ the position $p$ where the smaller segment strictly \\ containing $p$ is located.} & $O(\log(n))$\\ \\
Rmq($bv, l, r$) & \makecell[l]{Given a bit vector $bv$ and two positions $l$ and $r$,\\ returns the position $p$ where the node with \\minimal excess in the range $[l..r]$ is located.} & $O(\log(n))$\\ \\
PreOrderMap($bv, p$) &  \makecell[l]{ Given a bit vector $bv$ and a position $p$, returns \\ the index of the node in pre-order located in $p$.} & $O(1)$ \\ \\
PreOrderSelect($bv, i$) & \makecell[l]{Given a bit vector $bv$ and the index $i$ of the node  \\in pre-order traversal, returns the position in \\ the bit vector $bv$ where the node is located.} & $O(1)$\\ \\
PostOrderSelect($bv, i$) & \makecell[l]{Given a bit vector $bv$ and the index $i$ of the \\ node in post-order traversal, returns the position \\ in the bit vector $bv$ where the node is located.} &$O(\log(n))$\\ \\
FirstChild($bv, p$) & \makecell[l]{Given a bit vector $bv$ and a position $p$, returns \\ the position where the first child of $p$ is located.} & $O(1)$\\ \\
IsLeaf($bv, p$) & \makecell[l]{Given a bit vector $bv$ and a position $p$, returns \\ True if $p$ is a leaf and False otherwise.} & $O(1)$\\ \\
LCA($bv, l,r$) & \makecell[l]{Given a bit vector $bv$ and two positions $l$ and $r$, \\ returns the position where the lowest common \\ ancestor between $l$ and $r$ is located.} & $O(\log(n))$\\ \\
ClusterSize($bv, p$) & \makecell[l]{Given a bit vector $bv$ and a position $p$, returns \\ the size of the cluster where $p$ is the root.} & $O(\log(n))$\\ \\
NumLeaves($bv, p$) & \makecell[l]{Given a bit vector $bv$ and a position $p$, returns \\ the number of leaves that are below $p$.} & $O(\log(n))$\\ \\
\bottomrule
\end{tabular}
\end{table}

\subsection{Robinson-Foulds distance computation}
To compute the Robinson-Foulds distance, our algorithm receives as input two bit vectors that represent the two trees being compared, as well as a mapping between tree indexes. This mapping can represented by an array of $n$ integers such that the position of each integer corresponds to the index minus one on the first tree, and the integer value at that position in the array corresponds to the index on the second tree. This way it is possible to get the index of the node with a given label in the second tree given the index of where it is in the first tree. Notice that we only maintain the indexes that correspond to have labels, the others remain with value zero. An example of this map for the trees in Figures~\ref{fig:tree1} and~\ref{fig:tree2} is
\begin{equation}\label{vector}
\begin{array}{rcccccccccccc}
\text{index on $T_5$} & [ & 0 & 0 & 0 & 9 & 7 & 10 & 0 & 3 & 4 & 5 & ]\\
  \text{index on $T_4$ (-1)} &   & 0 & 1 & 2 & 3 & 4 & 5  & 6 & 7 & 8 & 9 & \\
\end{array}.
\end{equation}
As mentioned before, this mapping can be represented in a more compact form.
If we take the labels in the trees and enumerate them in some canonical order, assigning them identifiers from $1$ to $n$, then each tree represents a permutation $\pi$ of that enumeration.
Hence, given $T$ and the associated permutation $\pi$, $\pi[i]$ is the label at the node of $T$ with index $i$, and $\pi^{-1}[i]$ is the index of the node of $T$ with label $i$, for $i\in\{1,\ldots,n\}$.
By composing both permutations we can determine the bijection among the nodes of both trees.
Since each permutation can be represented with $(1 + \varepsilon) n\log n$ bits, allowing to find $\pi[i]$ in constant time and $\pi^{-1}[i]$ in $O(1/\varepsilon)$ time~\cite{navarro2016compact}, we obtain a more compact representation than the previous construction.
In what follows, for simplicity, we consider the mapping representation through an array of integers.
\begin{figure}[H]
  \centering
    \subcaptionbox{A rooted labelled phylogenetic tree $T_5$  with labels on leaves. $u_1$, $u_2$, $u_3$ and $u_4$ are unlabelled nodes.}{
  \scalebox{1}{
\begin{tikzpicture}
			\node (d) at (1,0) {D};
			\node (e) at (2,0) {E};
			\node (f) at (3,0) {F};
			\node (b) at (4,0) {B};
			\node (a) at (5,0) {A};
            \node (c) at (6,0) {C};
			\node (def) at (2,2) {$u_1$};
			\node (ac) at (5.5,1) {$u_2$};
			\node (abc) at (5,2) {$u_3$};
			\node (all) at (3.5,4) {$u_4$};
 		\draw  (d) |- (def);
			\draw  (e) |- (def);
			\draw  (f) |- (def);
			\draw  (a) |- (ac);
			\draw  (c) |- (ac);
            \draw  (b) |- (abc);
			\draw  (ac) |- (abc);
			\draw  (def) |- (all);
			\draw  (abc) |- (all);
                \end{tikzpicture}
             }}
  \hfill
      \subcaptionbox{Index $i$ of each node in $T_5$.}{
  \scalebox{0.8}{
  \begin{forest}
   for tree={
        grow=south,
         minimum size=5ex, inner sep=1pt,
        s sep=7mm
            }
    [1,circle,draw
        [2,circle,draw
            [3,circle,draw=blue, top color=blue!5, bottom color=blue!30
                [\textbf{D}, no edge]
            ]
            [4,circle,draw=orange, top color=orange!5, bottom color=orange!30
                [\textbf{E}, no edge]
            ]
            [5,circle,draw=brown, top color=brown!5, bottom color=brown!30
                [\textbf{F}, no edge]
            ]
        ]
        [6,circle,draw, name=cluster
            [7,circle,draw=red, top color=red!5, bottom color=red!30
                [\textbf{B}, no edge, name=lowestl]
            ]
            [8,circle,draw
                [9,circle,draw=green, top color=green!5, bottom color=green!30
                    [\textbf{A}, no edge]
                ]
                [10,circle,draw=yellow, top color=yellow!5, bottom color=yellow!30
                    [\textbf{C}, no edge,name=lowestr]
                ]
            ]
        ]
    ]
    \node[draw,fit={(cluster) (lowestl) (lowestr)}] {};
  \end{forest}
}}

\vspace{3ex}

 \subcaptionbox{The balanced representation of $T_5$.}{  
$
B =
\begin{array}{cccccccccccccccccccc}
 ( & ( & ( & ) & ( & ) & ( & ) & ) & ( & ( & ) & ( & ( & ) & ( & ) & ) & ) & )\\
1_o & 2_o & 3_o & 3_c & 4_o & 4_c & 5_o & 5_c & 2_c & 6_o & 7_o & 7_c & 8_o & 9_o & 9_c & 10_o & 10_c & 8_c & 6_c & 1_c\\
\end{array}
$}
  \caption{Tree $T_5$ and its balanced parentheses representation. Each pair of parentheses corresponds to a unique node \texttt{index} identified as $i$ in the tree.}\label{fig:tree2}
\end{figure}

Then the algorithm counts the number of clusters present in both trees. This is achieved by going through the first tree in a post-order traversal  and verifying, for all clusters, if they are present in the second tree.
To verify if a cluster from the first tree is present in the second tree, the key idea is to obtain the cluster from the second tree that is described by the lowest common ancestor (LCA) among all the taxa present in that cluster. Then if the number of taxa present in the cluster that was obtained is equal, we can guarantee that the cluster is present in both trees.
For example, consider $T_4$ and $T_5$ as the trees represented in Figures~\ref{fig:tree1} and \ref{fig:tree2} respectively.
To verify if the cluster highlighted in $T_4$, namely, $\{A, B\}$ is present in $T_5$, we need first to calculate the indexation of both taxa $A$ and $B$. In the balanced parentheses representation of $T_4$, $A$ and $B$ occurs in positions $4$ and $5$, respectively. Then, using the map, we can conclude that in balanced parentheses representation of $T_5$, $A$ and $B$ occurs in positions $9$ and $7$, respectively. Then by computing the LCA in $T_5$ between those nodes, it is obtained the index $6$, which represents the cluster highlighted in $T_5$. Given that, this cluster contains three leafs and the original one just two, and thus it can be concluded that cluster $A, B$ is not present in $T_5$. 

To efficiently compute this distance, the trees are traversed in a post-order traversal. This is done to
guarantee that all nodes are accessed before their ancestors. This way it is possible to minimize the
number of times that the \texttt{LCA} operation is called. For example, considering $T_4$ as the tree represented
in Figure~\ref{fig:tree1}. As mentioned before, when computing the LCA between the nodes $A$ and $B$, which value is $6$, we can conclude that the cluster is not present on tree $T_5$. When computed, each LCA value is saved on a stack associated to the node with index $3$ of $T_4$. 

Continuing the post-order, we know obtain for taxa $C$ that its position in $T_4$ is $6$ and, consulting the map, its position in $T_5$ is $10$.
Then, to check if the cluster ${A,B,C}$ is also present in $T_5$ we use the LCA stored in node with index $3$ of $T_4$, i.e., $6$ and position $10$ to compute the new LCA value. The new LCA value is also position $6$
and is saved on a stack associated to the node with index $2$ of $T_4$. Since in this case the cluster obtained in $T_5$ has the same number of leaves as the cluster ${A,B,C}$ in $T_4$ we can conclude, by construction that is the same cluster. This process continues until it reaches the tree root. Notice that when there are only labels on the leaves, the internal nodes are not included in clusters. 

In our approach we do not count the singleton clusters, since their occur in both trees. Moreover, the internal nodes of each tree give us the number of non singleton clusters present on that tree.
Then, knowing the number of clusters from both trees and the number of clusters that are present in both trees, we can conclude how many clusters are not common to both trees, therefore knowing the Robinson-Foulds distance. In Section~\ref{Implementation}, we present more details on the implementation of RF algorithm.

\subsection{The extended/weighted Robinson-Foulds}
Our algorithm can also compute the Robinson-Foulds distance for trees with taxa in internal nodes and/or with weights on edges (eRF, wRF, weRF). For internal labelled nodes, the approach is very similar to the previous one. The only difference is that it is also necessary to compute the LCA for the taxa inside the internal nodes. Then, to compare if the clusters are the same, instead of comparing the number of leaves, we compare if the size of the clusters are the same.  More details on these differences can be found in Appendix~\ref{appendix1}. 

Computing the weighted Robinson-Foulds distance can be done by storing the total sum of the weights of both trees. Then, we verify if each cluster is present in the second tree; if that is the case we subtract the weights of the cluster in both trees and add the absolute difference between the two of them. This approach can be seen as first considering that all clusters are present in only one tree and then correcting for the clusters that are found in both trees. More details on these differences can be found in Appendix~\ref{appendix1}.

\subsection{Information about the clusters}\label{clusterinfo}
When applying the algorithms described above, it is also possible to store the clusters that are detected in both trees. This can be done through adding to a vector the indexes of the cluster in the first and second tree when concluding that they are the same. Doing this allows us to check if the distance returned by the algorithm is correct and if the differences between the trees is well represented by the distance computed.

\section{Implementation}\label{Implementation}

Our algorithms are implemented in C\texttt{++} and are available at \url{https://github.com/pedroparedesbranco/TreeDiff}. All algorithms are divided in two phases, namely the parsing phase and the distance computation.
The first phase of all implemented algorithms receives two trees in newick format. Then the goal is to parse this format, create the two bit vectors that represent both trees and the mapping (\texttt{CodeMap}) to correlate the indexes of both trees.
The second phase of each algorithm receives as input the two bit vectors and the mapping, and computes the corresponding distance.
To represent the bit vectors we used the Succinct Data Structure Library (SDSL)~\cite{gbmp2014sea} which contains three implementations to compute the fundamental operations mentioned in Section~\ref{subsec:operations}. We chose the \texttt{bp\_support\_sada.hpp}~\cite{10.1145/2601073} since it was the one that obtained the best results in terms of time and space requirements.
The mapping that correlates the taxa between both trees relies on integer keys of 32 bits in our implementation.

\subsection{Parsing phase}
In the first phase, when parsing the newick format, the construction of the bit vectors is straightforward since the newick format already has the parentheses in order to represent a balanced parentheses bit vector. The only detail that we need to take into account is that when we encounter a leaf, we need to add two parentheses, the first one open and the second one closed (see Algorithm~\ref{alg:parse}).
The mapping (\texttt{CodeMap}) exemplified in Equation~\ref{vector} is also built in the first phase. For this process, we need to have a temporary hash table that associates the labels found in the newick representation to the indexes of the first tree (\texttt{HashTable}).
When a labelled node is found while traversing the first tree, it is added to the \texttt{HashTable} associating the label and the index where it is located in the tree. Then, when a labelled node is found in the second tree, the algorithm looks for that label in the \texttt{HashTable} to find the index $i$ where it is located in the first tree. Then it stores in the position $i-1$ of the \texttt{CodeMap} the index where that label is located in the second tree.

To compute the \texttt{wRF} distance it is also necessary to create two float vectors to save the weights that correspond to a given
cluster $c$
for each tree, i.e. the $w_T(c)$ in Definition~\ref{def3}. These weights are stored in the position that corresponds to the index where they are located in each tree. This process can be seen in Algorithm~\ref{alg:parse}.
The first phase takes linear time with respect to the number of nodes if we have a weighed or fully labelled tree, otherwise it takes linear time with respect to labelled nodes.

\begin{algorithm}[H]
\begin{algorithmic}[1]
\State \textbf{Input}: {$T_1, T_2$} in newick format.
\State \textbf{Output}: {${bv_1, bv_2, w_1, w_2, CodeMap, weightsSum}$} \Comment{{\em weightsSum} is only used for wRF and eWRF distances}
\State $HashTable \gets null$
\State $weightsSum \gets 0$
\For{$i = 1; i < 3; i++$}
    \State $bv_i \gets null$
    \State $index \gets$ 0
    \State  $s \gets null$ \Comment{$s$ is a stack of integers}
    \While{$c \gets getChar(T_i) \neq ;$}
        \If{$c == ($}
            \State $push(index, s)$
            \State $index$++
            \State $push\_back(bv_i, 1)$
        \ElsIf{$c == ','$}
            \State continue
        \ElsIf{$c == )$}
            \State $c \gets getChar(T_i)$
            \If{ \textbf{not} ($c == ')' \textbf{ or }  c == '(' \textbf{ or } c == ',' \textbf{ or } c == ';'$)}
                \If{$c == :$}
                    \State $c \gets getChar()$
                    \State $weight \gets getWeight(T_i)$
                    \State $w_i[index] = weight$
                    \State $weightsSum = weightsSum + weight$
                \Else
                    \State $label \gets getLabel(T_i)$\;
                    \If{$i == 1$}
                        \State $HashTable[label] = index$
                    \Else
                        \State $CodeMap[HashTable[label]] = count$
                    \EndIf
                \EndIf
            \Else
                \State $c \gets ungetChar(T_i)$
            \EndIf
            \State $pop(s)$
            \State $push\_back(bv_i, 0)$
        \Else
            \State $push\_back(bv_i, 1)$\;
            \State $push\_back(bv_i, 0)$\;
            \State $label \gets getLabel(T_i)$\;
            \If{$i == 1$}
                \State $HashTable[label] = index$
            \Else
                \State $CodeMap[HashTable[label]] = count$
            \EndIf
        \EndIf
    \EndWhile
\EndFor 
\State $free(HashTable)$
\State \textbf{Return} ${bv_1, bv_2, w_1, w_2, CodeMap, weightsSum}$

\caption{Parsing phase}
\label{alg:parse}
\end{algorithmic}
\end{algorithm}

\subsection{Distance calculation}
The second phase is the computation of the Robinson-Foulds distance or one of its variants. We extended the functionality of SDSL to support more operations, as previously mentioned. More details are available in Appendix~\ref{appendix2}.

For the RF distance and its variants, it was implemented two different approaches that guarantees that the algorithm traverses the tree in a post-order traversal. The first one, which we designate by \texttt{rf\_postorder}, takes advantage of the \texttt{PostOrderSelect} operation while the second one, which we designate by \texttt{rf\_nextsibling}, takes advantage of the \texttt{NextSibling} and \texttt{FirstChild} operations.

\subsubsection{Robinson-Foulds using \texttt{PostOrderSelect}}

To make sure that the tree is traversed in a post-order traversal, this implementation simply does a loop from 1 to $n$ and calls the function \texttt{PreOrderSelect} for all the values. This implementation also uses a very similar strategy than the one used in Day algorithm~\cite{dayalgorithm} to keep the LCA results computed earlier.
This strategy consists in using a stack that keeps track, for each node, of the index of the corresponding cluster from the second tree and the size of the cluster from the first tree. Then, whenever it is found a leaf, it is added the corresponding index as well as the size of the leaf which is one. When the node found is not a leaf, it goes through the stack and finds the last nodes that where added until the sum of their sizes is equal to the size of the cluster. For these nodes it is computed the LCA between each pair of them while removing them from the stack. When this process is done it is possible to verify if that cluster is present in the other tree and then add the information of that cluster to the stack so that the computations of the LCA that where made are not done again. An implementation of this process can be seen in Algorithm~\ref{alg:treediff1}.
\begin{algorithm}[H]
\caption{\texttt{rf\_postorder} implementation\label{alg:treediff1}}
\begin{algorithmic}[1]
\State \textbf{Input}: {$bv_1, bv_2, CodeMap$}\\
\Comment{Let $numInternalNodes1$ and $numInternalNodes2$  be the number of internal nodes of both trees.}
\State \textbf{Output}: Robinson-Foulds distance
\\
\State $equalClusters \longleftarrow 0$\;
\For{$i \gets 1$ to $N$}
     \State $p \gets$ \textbf{PostOrderSelect}($bv_1, i$)\;
     \If{\textbf{IsLeaf}($bv_1, p$)}
        \State $lca \gets$ \textbf{PreOrderSelect}(CodeMap$[$\textbf{PreOrderMap}($bv_1, p$) - 1$]$ + 1)\; 
        \State $size \gets 1$\;
        \State push($s, <lca, size>$)\; \Comment{Let $s$ be a stack.}
     \Else
        \State $cs \gets$ \textbf{ClusterSize}($bv_1, p$) - 1\;
        \While{$cs \neq 0$}
            \State $<lca, size> \gets$ pop($s$)\;
            \If{$lcas != null$}
                \State $<lca, size> \gets$ pop($s$)\;
                \State $lcas \gets$ \textbf{lca}($bv_2, lcas, lca$)\;
            \EndIf
            \State $cs = cs - size$\;
        \EndWhile
        \If{\textbf{NumLeaves}($bv_1, p$) = \textbf{NumLeaves}($bv_2, lcas$)}
            \State $equalClusters \gets equalClusters$ + 1\;
        \EndIf
        \State push($s$,$<lcas,$ \textbf{ClusterSize}($bv_1, p$) + $1>$)\;     
        \State $lcas \gets$ null\;
     \EndIf
\EndFor
\State $distance \gets (numInternalNodes1 + numInternalNodes2 - equalClusters*2) / 2$\;
\State \Return distance\;
\end{algorithmic}
\end{algorithm}
Despite the strategy of using a stack is similar to Day algorithm, our approach just stores in the stack 2 integers for each node, while in Day algorithm it is necessary to store 4 integers for each node, namely: the value of the left leaf; the value of the right leaf; the number of leafs below; and the size of the cluster. Moreover, with our approach there is no need to create the clusters table as Day need since we use the LCA to verify the clusters.

Given that the algorithm goes through each node in the first tree and needs to compute the \texttt{LCA}, the \texttt{PostOrderIndex} and the number of leaves \texttt{NumLeaves} from each cluster, the theoretical complexity of the algorithm is $O(n\log(n))$, with $n$ the number of nodes in each tree.

\subsubsection{Robinson-Foulds using \texttt{NextSibling} and \texttt{FirstChild}}

This implementation takes a slightly different approach. It uses a recursive function that is called for the first time for the root node. Then it verifies if the given node is not a leaf and, if that is the case, it calls the function to the first child node. Then for all the calls it goes through all the siblings of the given node and computes the LCA between all of them. Whenever the node in question is not a leaf, the algorithm uses the LCA value computed and the operation \texttt{NumLeaves} to verify if the cluster is common to both trees. The function returns the index of the LCA obtained so that in this implementation there is no need to keep an explicit stack to reuse the LCA values computed throughout the algorithm. An implementation of this process can be seen in Algorithm~\ref{alg:treediff2}.

\begin{algorithm}[H]
\begin{algorithmic}[1]
\State \textbf{Input}: {$bv_1, bv_2, CodeMap$} \\
\Comment{To guarantee a post-order transversal, the initial call to this function initiates current to tree root index}\\
\Comment{Let $numInternalNodes1$ and $numInternalNodes2$  be the number of internal nodes of both trees.}
\State \textbf{Output}: Robinson-Foulds distance\\

\State  $equalClusters \longleftarrow 0$\;
\State $\textsc{\texttt{rf\_nextsibling\_aux}}(0)$\;
\State $distance \gets$ ($numInternalNodes1 + numInternalNodes2 - equalClusters*2) / 2$\;
\State \Return distance\;
\\
\Procedure{$\texttt{rf\_nextsibling\_aux}$}{$current$}: int

    \If{\textbf{IsLeaf}($v1, current$)}
        \State $lcas \gets$ \textbf{PreOrderSelect}(CodeMap$[$\textbf{PreOrderMap}($v1, current$) - 1$]$ + 1\;
    \Else
        \State $lcas \gets$ \textbf{rf\_nextsibling}(FirstChild($v1, current$))\;
        \If{\textbf{NumLeaves}($v1, current$) = \textbf{NumLeaves}($v2, lcas$)}
            \State $equalClusters \gets equalClusters$ + 1\;
        \EndIf
    \EndIf
    \While{\textbf{NextSibling}(v1, current)}
        \State current $\gets$ \textbf{NextSibling}($v1, current$)\;
        \If{\textbf{IsLeaf}($v1, current$)}
        \Else
            \State lcas\_aux $\gets$ \textbf{rf\_nextsibling}(FirstChild($v1, current$))\;
            \State $lcas \gets$ \textbf{lca}(v2, $lcas, lcas\_aux$)\;
            \If{\textbf{NumLeaves}($v1, current$) = \textbf{NumLeaves}($v2, lcas$)}
                \State $equalClusters \gets equalClusters$ + 1\;
            \EndIf
        \EndIf
    \EndWhile
    \State \Return $lcas$\;
\EndProcedure
\end{algorithmic}
\caption{\texttt{rf\_nextsibling} implementation}
\label{alg:treediff2}
\end{algorithm}

\subsubsection{Weighted/extended Robinson-Foulds}
We have also implemented eRF, wRF and weRF distances using \texttt{PostOrderSelect}, and using \texttt{NextSibling} and \texttt{FirstChild}. There are just a few differences with respect to Algorithms~\ref{alg:treediff1} and~\ref{alg:treediff2}. For instance, when phylogenetic trees are fully labelled, it is also necessary to take into account the labels of internal nodes for the LCA computation. In the weighted variations, it is also necessary to have two vectors of size $n$, where $n$ is the number of nodes of each tree, and the variable \texttt{weightsSum}, which is updated during the tree transversal. More details can be found in Appendix~\ref{appendix2}, with the implementation of eRF and wRF using \texttt{PostOrderSelect}.

\subsection{Time and memory analysis}
We also implemented the Day algorithm so that we could compare it with our implementation in terms of running time and memory usage. This algorithm was also implemented in C\texttt{++} and stores two vectors of integers with size $n$ for each tree and two vectors of integers with size $n/2$ (the number of leaves). The complexity of the algorithm is $O(n)$ with respect to both time and space.

The complexity of both parsing phases, namely the Day approach and our approach, are $O(n)$ since they loop through all the nodes. The complexity of the  Day algorithm with respect to the RF computation phase, it is also $O(n)$. With respect to the second phase of our approach, we will depict the complexity analysis with respect to the number of times the \texttt{LCA}, \texttt{NumLeaves}, \texttt{ClusterSize}, \texttt{PostOrderSelect} and \texttt{NextSibling} are called and which depends on $n$.

For the RF computed with the \texttt{rf\_postorder} implementation, the \texttt{PostOrderSelect} operation is called for each node $n$ times. The number of times the \texttt{LCA} is called is equivalent to the number of leafs minus the number of first leafs which in the worst case is $n - 2$. This case would be a tree that only contains the root as an internal node. With respect to the \texttt{NumLeaves} operation, it is called two times for each internal node, with the worst case being $2 \times (n - 1)$. This case is when a tree only has one leaf. The \texttt{ClusterSize} operation is called one time for each internal node which would result in $n - 1$ in the worst case as before. In total, the complexity would be $(n + (n - 2) + (2 \times (n - 1)) + (n - 1)) \times O(\log(n)) = (5n - 5) \times O(\log(n))$. However, the worst case for the \texttt{LCA} operations is the best case for the \texttt{NumLeaves} and \texttt{ClusterSize} and vice-versa. This means that in the worst case, the \texttt{LCA} operation would be applied 0 times and the complexity would be $(4n - 3) \times O(\log(n)))$.

For the RF computed with the \texttt{rf\_nextsibling} implementation, the number of calls for the \texttt{LCA} and \texttt{NumLeaves} are the same as before. The number of \texttt{NextSibling} operation calls is the same as the \texttt{LCA} operation. This means that the complexity for this case is $((n - 2) + (n - 2) + (2 \times (n - 1))) \times O(\log(n)) = (4n - 5) \times O(\log(n))$, however for the same reason as before it can be reduced to $(2n - 2) \times O(\log(n))$. 

For the wRF distance computed by both the \texttt{rf\_postorder} and \texttt{rf\_nextsibling} implementations, the number of times these operations are called is the same as the original RF.
For the eRF and weRF computed by both implementations the only difference is that the \texttt{LCA} operation is called also for all internal nodes. This means that the number of times the \texttt{LCA} operation is applied is equivalent to the total number of nodes minus the number of first leafs. This would change the complexity in both of the implementations. In \texttt{rf\_postorder} it would change to $(5n - 4) \times O(\log(n))$ since the number of \texttt{LCA} calls passed from 0 to $n - 1$ in the worst case. In \texttt{rf\_nextsibling} it would change to $(3n - 3) \times O(\log(n))$. All implementations referred above have then a running time complexity of $O(n \log(n))$.

In terms of memory usage, the RF computed by both the \texttt{rf\_nextsibling} and \texttt{rf\_postorder} implementations only uses two bit vectors of size $2n$ and a vector of 32bit integers with size $n$ so the total of bits used is $36n$ bits, however the last one needs an additional stack to save the LCA values. And the efficient implementation of primite operations over the bit vectors require extra $o(n)$ bits. The Day algorithm (\texttt{rf\_day} implementation) needs four vectors of 32bit integers with size $n$ as well as two vectors of 32bit integers with the size equal to the number of leafs which is equal to $n - 1$ in the worst case. This results in $32 \times 4n + 32 \times 2(n - 1) = 192n - 64$ bits.

\section{Experimental evaluation and discussion}
  In this section it is discussed the performance of the \texttt{rf\_postorder} and \texttt{rf\_nextsibling} implementations in comparison to their baselines and extensions. To evaluate the implementations we used the ETE Python library to create a random generator of trees in newick format. To evaluate the RF distance, it was generated $10$ trees ($5$ pairs) starting with $10,000$ leaves and going up to trees with $100,000$ leaves, with a step of $10,000$. It was generated not only five pairs of trees with information only on the leaves for all the sizes tested, but also for fully labelled trees. To analyse the memory usage it was used the valgrind massif tool~\cite{nethercote2003valgrind}.

\begin{figure}[H]
\centering
\includegraphics[width=1.0\textwidth]{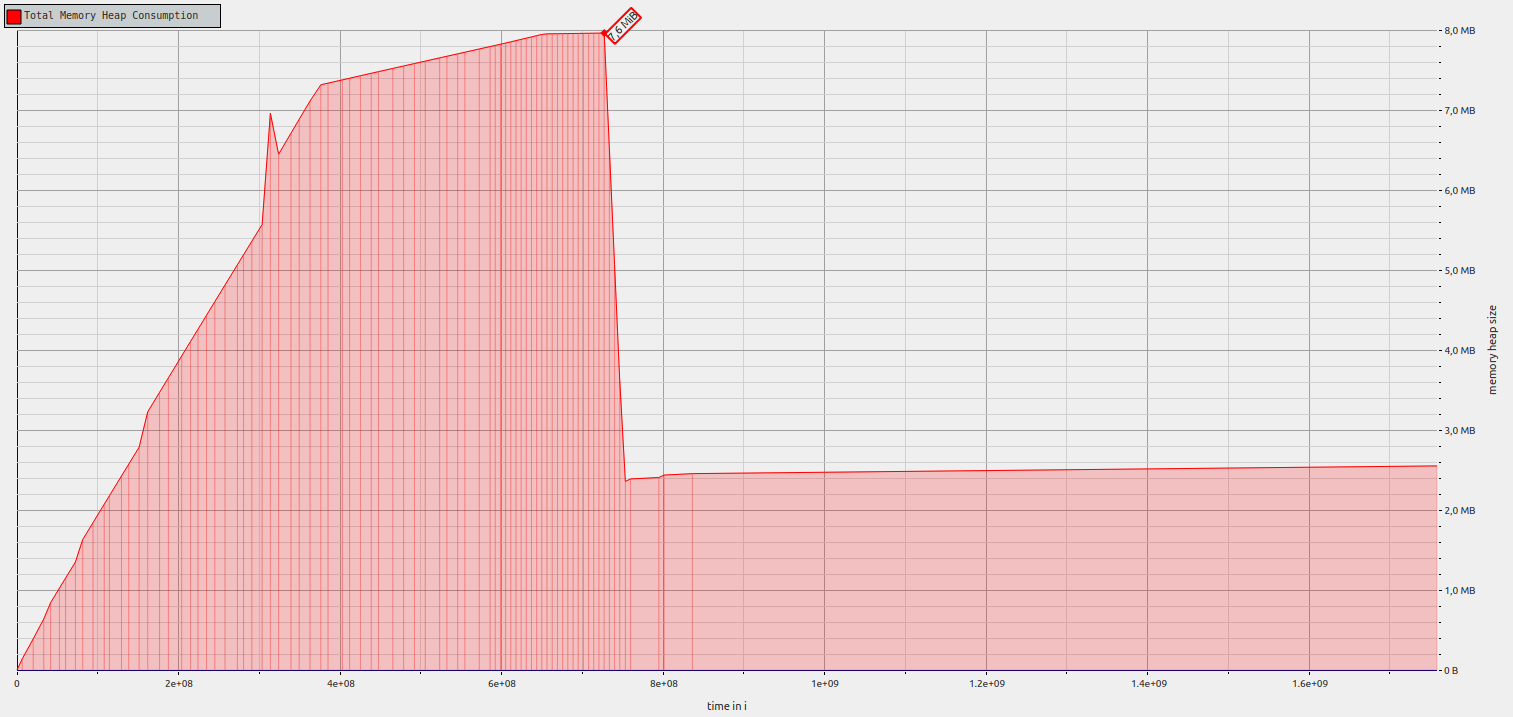}
\caption{Heap allocation profile for two trees with 100000 leaves in \texttt{rf\_postorder} implementation.}
\label{fig:memoryCharts}
\end{figure}

First it was analysed the memory used throughout the \texttt{rf\_postorder} algorithm execution for a tree that with $100,000$ leaves. Figure~\ref{fig:memoryCharts} shows that for the first phase, the memory consumption is higher since a hash table has to be used to create the map that correlates labels and tree indexes. In the transition to the second phase, since the hash table is no longer needed and, as such, the memory it required is freed, the memory consumption falls abruptly. In the second phase of the algorithm, the memory consumption is lower since this phase only uses the information stored in the two bit vectors and related data structures, and in the mapping that correlates the two trees.
Note that by serializing and storing the trees and related permutations as their bit vector representations, we can avoid the memory required for parsing the newick format.

\begin{figure}[H]
\centering
\includegraphics[width=0.8\textwidth]{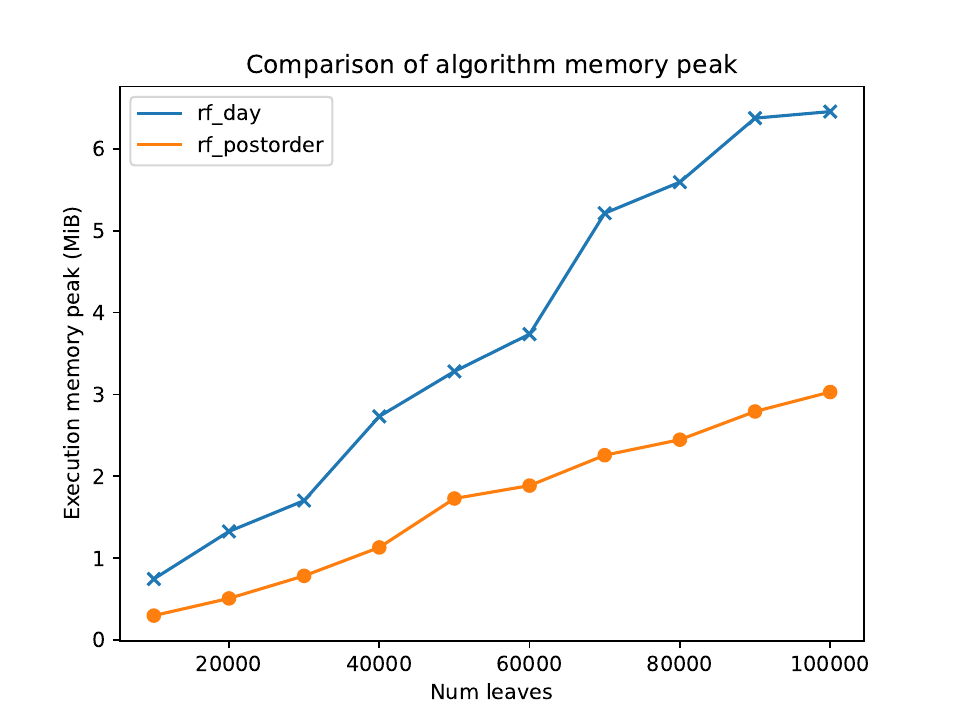}
\caption{Memory usage peak comparison.}
\label{fig:memoryComparison}
\end{figure}

Secondly, it was compared the memory usage for the second phase of the \texttt{rf\_postorder} and \texttt{rf\_day} implementations. Figure~\ref{fig:memoryComparison} shows the comparison between both algorithms in terms of their memory peak usage. The \texttt{rf\_postorder} exhibits a significant lower memory peak compared to the \texttt{rf\_day} algorithm.
  Note that the difference is smaller than expected in our theoretical analysis since in our implementations we are keeping track of common clusters, as described in Section~\ref{clusterinfo}, and extra space is required for supporting fast primitive operations over the bit vectors. If we omit that cluster tracking mechanism, we can reduce further the memory usage of our implementations. 

\begin{figure}[H]
       \subcaptionbox{Algorithm phase.\label{fig:approachphase}}{\includegraphics[width=0.5\textwidth]{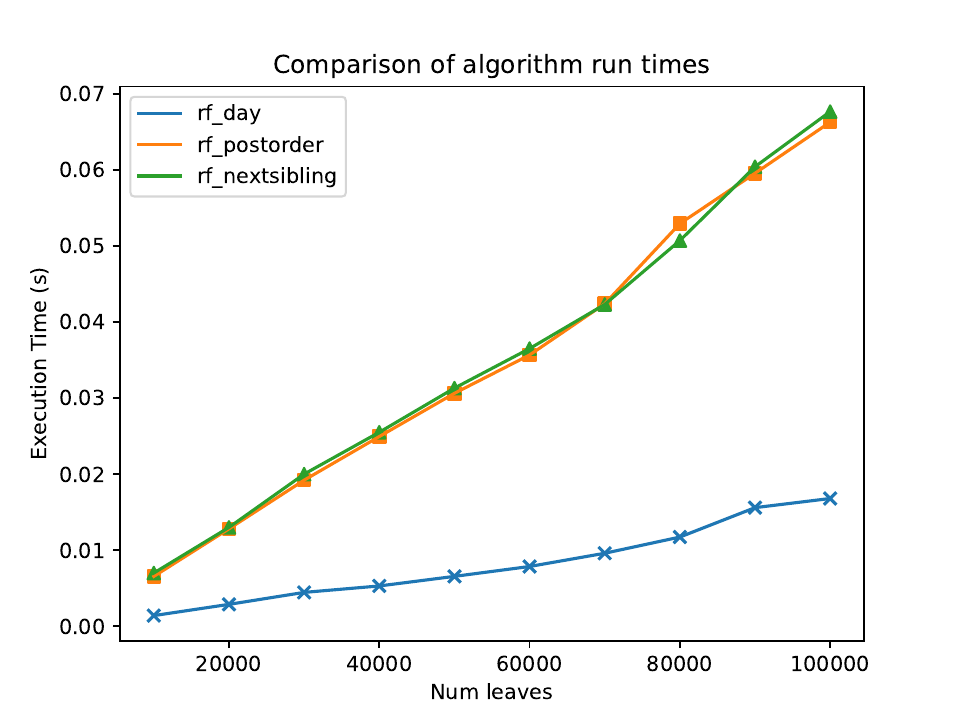}}  
       \subcaptionbox{Parsing phase.\label{fig:parsingphase}}{\includegraphics[width=0.5\textwidth]{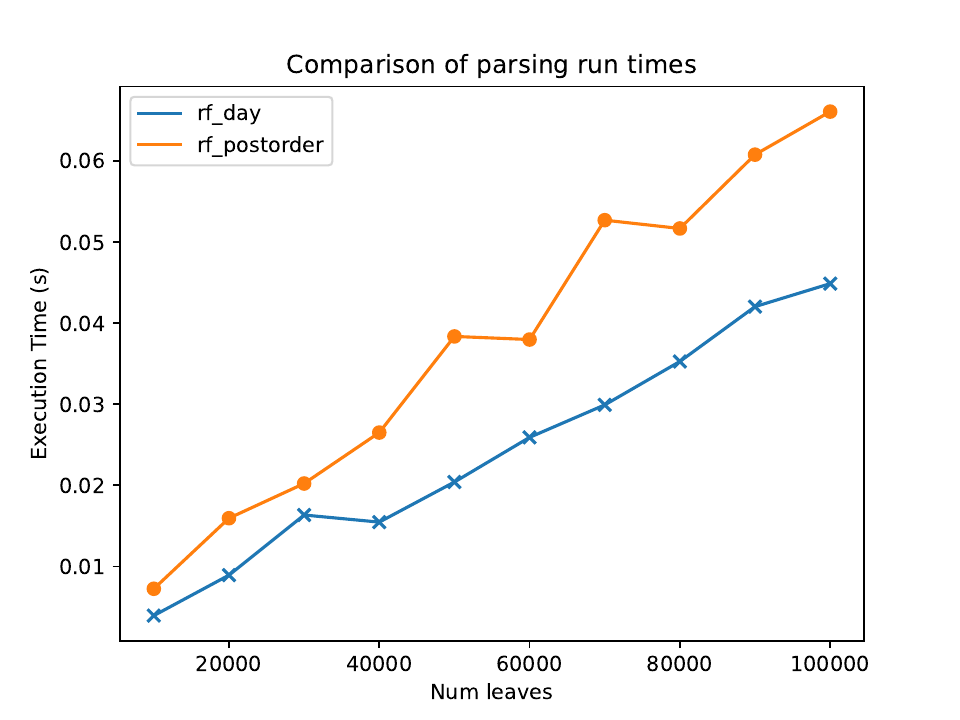}}   
        \caption{Run time for trees with different sizes.}
        \label{fig:runTimeComparison}
\end{figure}

Then, it was compared the running time between the two algorithms for the parse and algorithm phases: Figure~\ref{fig:runTimeComparison}. For the parse phase, we
only compare the \texttt{rf\_postorder} with the \texttt{rf\_day} since the \texttt{rf\_nextsibling} has the exact same parsing phase as \texttt{rf\_postorder}. It can be observed in Figure~\ref{fig:parsingphase} that the differences between the running time of the two implementations are not significant, indicating similar performances between the two algorithms. In the second phase, the comparison was made between all three implementations. In Figure~\ref{fig:approachphase}, it can be observed that both the \texttt{rf\_postorder} and \texttt{rf\_nextsibling} implementations presented almost the same results. Even though the theoretical complexity of those implementations is $O(n\log(n))$, in practise it seems to be almost linear. This finding suggests that the operations used to traverse the tree tend to be almost $O(1)$ in practise even though they have a theoretical complexity of $O(\log(n))$. The difference in the running time between these implementation and the \texttt{rf\_day} one can be explained by the number of operations that are computed for each node as explained in previously, and because we are using succinct tree representations.

Next, it was compared the running time for the RF and eRF distances. Note that the Day algorithm does not support fully labelled trees. In both of the \texttt{rf\_postorder} and \texttt{rf\_nextsibling} implementations, the eRF distance seems to be also linear in practise. However, for the \texttt{rf\_nextsibling} implementation, the running time difference seems to be much insignificant than the difference observed for the \texttt{rf\_postorder} implementation. This results are consistent with the theoretical analysis since in the \texttt{rf\_nextsibling} implementation the number of times each operation is applied in the worst case is $(3n - 3)$ and in the \texttt{rf\_postorder} is $(5n - 4)$. This concludes that the \texttt{rf\_nextsibling} implementation gives better results for most trees when it is used to compute the distance for fully labelled trees.
\begin{figure}[H]
  \subcaptionbox{Using \texttt{NextSibling} and \texttt{FirstChild}.}{\label{fig:nextsiblinglabelled}\includegraphics[width=0.5\textwidth]{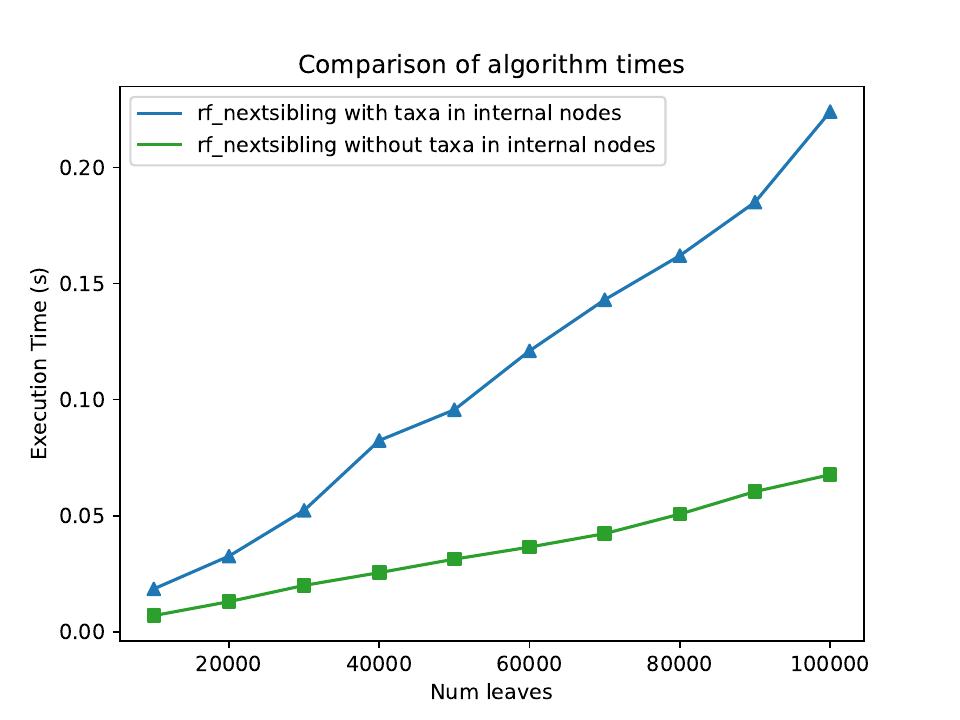}}  
       \subcaptionbox{Using \texttt{PostOrderSelect}.}{\label{fig:postorderlabelled}\includegraphics[width=0.5\textwidth]{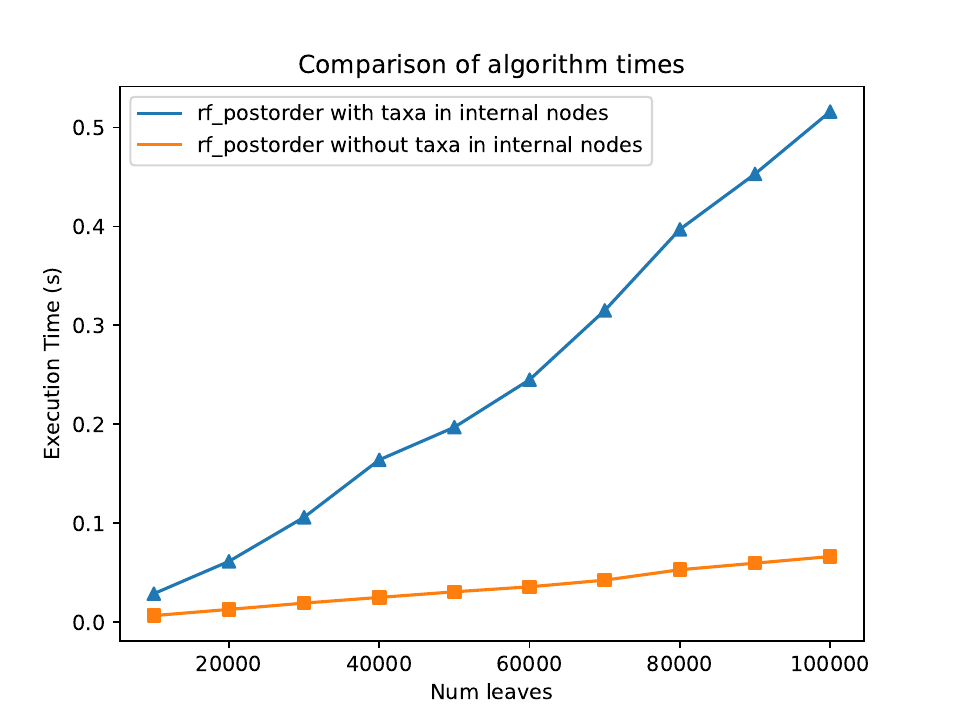}}   
        \caption{Run time comparison for leaf labelled trees and fully labelled trees.}
        \label{fig:rfLabelledTrees}
\end{figure}

As expected, our implementation is slower than the Day algorithm implementation, because we have the overhead of using a succinct representation for trees. But it is noteworthy that our implementation offers a significant advantage in terms of memory usage, namely when trees are succinctly serialized and stored. This trade-off between memory usage and running time is common in this setting, with succinct data structures being of practical interest when dealing with large data.

\section{Conclusions}
We provided implementations of the Robinson-Foulds distance over trees represented succinctly, as well as its extension to fully labelled and weighted trees.
These kind of implementations are becoming useful as pathogen databases increase in size to hundreds of thousands of strains, as is the case of EnteroBase~\cite{zhou2020enterobase}.
Phylogenetic studies of these datasets imply then the comparison of very large phylogenetic trees.

Our implementations run theoretically in $O(n\log n)$ time and require $O(n)$ space.
Our experimental results show that in practise our implementations run in almost linear time, and that they require much less space than typical implementations.
The use of succinct data structures introduce a slowdown in the running time, but that is an expected trade-off; we gain in our ability to process much larger trees using space efficiently.

Each tree with $n$ nodes can be represented in $2n + o(n) + (1 + \varepsilon) n\log n$ bits, storing its balanced parentheses representation and its corresponding permutation through bit vectors.
Our implementations rely on and extend SDSL~\cite{gbmp2014sea}, making use of provided bit vector representations and primitive operations.
In our implementations we did not explore however the compact representation of the permutation associated to each tree.
Hence, the space used by our implementations can be further improved, in particular if we rely on that compact representation
to store phylogenetic trees on secondary storage.
We leave this as future work noting that techniques to represent compressible permutations may be exploited in this setting, leading to a more compact representation~\cite{navarro2016compact};
phylogenetic trees tend to be locally conserved and, hence, that favors the occurrence of smaller and larger runs in tree permutations.


\authorcontributions{Conceptualization, APF and CV; methodology, APF and CV; software, APB; validation, APB and CV; formal analysis, APB and APF; 
writing---original draft preparation, CV and APB; writing---review and editing, APF and CV 
}

\funding{The work reported in this article received funding from Funda\c{c}\~ao para a Ci\^encia e a Tecnologia (FCT) with references UIDB/50021/2020 (DOI:10.54499/UIDB/50021/2020), and from European Union’s Horizon 2020 research and innovation programme under Grant Agreement No. 951970 (OLISSIPO project). It was  also supported through Instituto Politécnico de Lisboa with project IPL/IDI\&CA2023/PhyloLearn\_ISEL.}

\appendixtitles{no} 
\appendixstart
\appendix
\section[\appendixname~\thesection]{}\label{appendix2}
The succint data structures library already provided implementation for most of the used operations  on \texttt{bp.support.sada.hpp} file. We extended this library to support other operations needed to compute the distances, namely \texttt{PreOrderMap}, \texttt{PreOrderSelect}, \texttt{PostOrderSelect}, \texttt{IsLeaf}, \texttt{LCA}, \texttt{ClusterSize} and \texttt{NumLeaves} and their implementation can be seen below. Moreover, it was added two operations that were already in the SDSL library but not present in the  \texttt{bp.support.sada.hpp} file. These operations were the \texttt{rank10} operation to enable us to count the number of leaves, and the \texttt{select0} operation to enable us to go through a tree in a post-order traversal.
\begin{algorithmic}[0]
\Procedure{$\texttt{PreOrderMap}$}{$bv, p$}: int
\State \hspace{0.5cm} \textbf{return} rank1($bv, p$)
\EndProcedure
\vspace{0.1cm}
\Procedure{$\texttt{PreOrderSelect}$}{$bv, i$}: int
\State \hspace{0.5cm} \textbf{return} select1($bv, i$)
\EndProcedure
\vspace{0.1cm}
\Procedure{$\texttt{PostOrderSelect}$}{$bv, i$}: int
\State \hspace{0.5cm} \textbf{return} findOpen(Select0($bv, i$))
\EndProcedure
\vspace{0.1cm}
\Procedure{$\texttt{IsLeaf}$}{$bv, i$}: int
\State \hspace{0.5cm} \textbf{return} bv$[i+1] == 0 $
\EndProcedure
\vspace{0.1cm}
\Procedure{$\texttt{LCA}$}{$bv, l, r$}: int
\State \hspace{0.5cm} \textbf{if} l > r:
    \State \hspace{1cm} l $\leftrightarrow$ r
    \State \hspace{0.5cm} \textbf{return} enclose($bv$,rmq($bv, l, r$) + 1)
\EndProcedure
\vspace{0.1cm}    
\Procedure{$\texttt{ClusterSize}$}{$bv, p$}: int
\State \hspace{0.5cm} \textbf{return} (findClose($bv, p$) - v + 1) / 2
\EndProcedure
\vspace{0.1cm}   
\Procedure{$\texttt{NumLeaves}$}{$bv, p$}: int
\State \hspace{0.5cm} \textbf{return} rank10($bv$, findClose($bv, p$)) - rank10($bv, p$) + 1
\EndProcedure
\vspace{0.1cm} 
\end{algorithmic}

\section[\appendixname~\thesection]{}\label{appendix1}

The algorithms presented in this appendix depict the RF variations, namely the extended Robinson-Foulds distance (eRF) and the weighted Robinson-Foulds distance (wRF).

\begin{algorithm}[H]\label{alg:treediff5}
\caption{Extended Robinson-Foulds (eRF)}
\begin{algorithmic}[1]
\State \textbf{Input}: {$bv_1, bv_2, CodeMap$}
\State \textbf{Output}: Robinson-Foulds distance for fully labelled trees
\\
\State $equalClusters \gets$ 0
\For{$i \gets 1$ to $N$} 
    \State $p \gets$ PostOrderSelect($v1, i$) 
    \State $lca \gets$ PreOrderSelect(CodeMap$[$PreOrderMap($v1, p$) - 1$]$ + 1) 
    \State $size \gets 1$
    \State push($<lca, size>, s$)
    \If{!IsLeaf($p$))}
        \State $cs \gets$ ClusterSize($v1, p$)
        \While{$cs \neq 0$}
            \If{$lcas = null$}
                \State $<lcas, size> \gets$ pop($s$)
            \Else
                \State $<lca, size> \gets$ pop($s$)
                \State $lcas \gets$ LCA($v2, lcas, lca$)
            \EndIf
            \State $cs = cs - size$
        \EndWhile
        \If{ClusterSize($v1, p$) = ClusterSize($v2, lcas$)}
            \State $equalClusters \gets equalClusters$ + 1
        \EndIf
        \State push($<lcas, ClusterSize($v1, p$)>, s$)
        \State $lcas \gets$ null
    \EndIf
\EndFor
\State $distance \gets$ (numInternalNodesv1 $+$ numInternalNodesv2 - equalClusters*2) / 2 \\
\Return distance
\end{algorithmic}
\end{algorithm}

\begin{algorithm}[H]\label{alg:treediff6}
\caption{weighted Robinson-Foulds (wRF)}
\begin{algorithmic}[1]
\State \textbf{Input}: {$bv_1, bv_2, CodeMap, w1, w2, weightsSum$}
\State \textbf{Output}: weighted Robinson-Foulds distance
\\
\State $equalClusters \gets$ 0
\For{$i \gets 1$ to $N$} 
    \State $p \gets$ PostOrderSelect($v1, i$) 
    \If{IsLeaf($p$))}
        \State $lca \gets$ PreOrderSelect(CodeMap$[$PreOrderMap($v1, p$) - 1$]$ + 1) 
        \State $size \gets 1$
        \State $weight1 \gets w1[PreOrderMap(v1, p) - 1]$
        \State $weight2 \gets w2[PreOrderMap(v2, lca) - 1]$
        \State $weightsSum = weightsSum - (weight1 + weight2 - abs(weight1 - weight2))$
        \State push($<lca, size>, s$)
    \Else
        \State $cs \gets$ ClusterSize($v1, p$) - 1
        \While{$cs \neq 0$}
            \State $<lca, size> \gets$ pop($s$) 
            \State $lcas \gets$ LCA($v2, lca, $)
            \If{$lcas = null$}
                \State $<lcas, size> \gets$ pop($s$)
            \Else
                \State $<lca, size> \gets$ pop($s$)
                \State $lcas \gets$ LCA($v2, lcas, lca$)
            \EndIf
            \State $cs = cs - size$
        \EndWhile
        \If{NumLeaves($v1, p$) = NumLeaves($v2, lcas$)}
            \State $weight1 \gets w1[PreOrderMap(v1, p) - 1]$
            \State $weight2 \gets w2[PreOrderMap(v2, lcas) - 1]$
            \State $weightsSum = weightsSum - (weight1 + weight2 - abs(weight1 - weight2))$
        \EndIf
        \State push($<lcas, ClusterSize($v1, p$)>, s$)
        \State $lcas \gets$ null
    \EndIf
\EndFor
\State \textbf{return} $weightsSum$
\end{algorithmic}
\end{algorithm}

\begin{adjustwidth}{-\extralength}{0cm}

\reftitle{References}
\bibliography{myBibliography}

\end{adjustwidth}
\end{document}